\documentclass[reprint,author-numerical]{revtex4-1}

\usepackage{graphicx}% Include figure files
\usepackage{dcolumn}% Align table columns on decimal point
\usepackage{bm}% bold math
\newtheorem{lemma}{Lemma}
\newtheorem{proposition}{Proposition}
\newtheorem{theorem}{Theorem}

\def\tr{\mathop{\rm tr}\nolimits}

\def\dif{\mathop{\rm d\!}\nolimits}

\newcommand{\be}{\begin{equation}}
\newcommand{\ee}{\end{equation}}

\begin{document}

%\preprint{AIP/123-QED}

\title[Intrinsic, deductive, explicit, and algorithmic  characterization of the Szekeres-Szafron solutions]{Intrinsic, deductive, explicit, and algorithmic characterization of the Szekeres-Szafron solutions}% Force line breaks with \\
%\thanks{Footnote to title of article.}

\author{Joan Josep Ferrando}
\altaffiliation[Also at ]{Observatori Astron\`omic, Universitat
de Val\`encia, \\E-46980 Paterna, Val\`encia, Spain}%Lines break automatically or can be forced with \\
 \email{joan.ferrando@uv.es.}
\affiliation{ Departament d'Astronomia i Astrof\'{\i}sica,
Universitat
de Val\`encia, E-46100 Burjassot, Val\`encia, Spain.}%\\This line break forced with \textbackslash\textbackslash

\author{Juan Antonio S\'aez}
\affiliation{ Departament de Matem\`atiques per a l'Economia i
l'Empresa,
Universitat de Val\`encia, E-46071 Val\`encia, Spain.%\\This line break forced% with \\
}%

\date{\today}% It is always \today, today,

\begin{abstract}
We write the known invariant definition of the Szekeres-Szafron
family of solutions in an intrinsic, deductive, explicit and
algorithmic form. We also intrinsically characterize the two
commonly considered subfamilies, and analyze other subclasses, also
defined by first-order differential conditions. Furthermore, we
present a Rainich-like approach to these metrics.

\end{abstract}

%Use showkeys class option if keyword
\maketitle

\section{\label{sec-intro}Introduction}

The Szekeres-Szafron family of solutions is known as one of the most
significant inhomogeneous cosmological models \cite{Krasinski,
Krasinski-Plebanski, EMM}. These metrics were obtained by Szafron
\cite{Szafron} as the generalization for a nonvanishing pressure of
the Szekeres dust solutions \cite{Szekeres}. The symmetries and
other physical and geometric properties of the Szekeres dust models
have been widely studied \cite{Krasinski, Krasinski-Plebanski,
bonnor, Bonnor-ST, berger, goode} (see also the recent papers
\cite{hellaby,G-hellaby} and references therein).

On the other hand, Szafron \cite{Szafron} also extended the
invariant characterization by Wainwright \cite{Wainwright} of the
Szekeres solutions to the nonvanishing pressure models:
\ \\[2mm]
{\bf Invariant characterization 1} \cite{Szafron}: A
Szekeres-Szafron metric is characterized by the following
conditions: (i) it is a perfect fluid solution with a geodesic and
irrotational unit velocity; (ii) the Weyl tensor is Petrov-Bel type
D and the velocity of the fluid lies in the two-plane $\Pi$ spanned
by the two null principal directions; (iii) any vector in the
two-plane $\Pi^{\perp}$ is an eigenvector of the shear; (iv) the
two-plane $\Pi$ admits orthogonal two-surfaces.

Note that the above statement (i) is a first-order differential
condition for the Ricci tensor. The second one is algebraic for both
the Ricci and the Weyl tensors. The third one is algebraic for the
Weyl tensor and a first-order differential condition for the Ricci
tensor. And (iv) is a first-order differential condition for the
Weyl tensor. So, characterization 1 imposes first-order differential
conditions on the Ricci and Weyl tensors.

In a later paper Szafron and Collins \cite{Szafron-Collins} offered
an alternative invariant characterization following from their study
of the inhomogeneous cosmological models with intrinsic symmetries
\cite{Collins-Szafron}.
\ \\[2mm]
{\bf Invariant characterization 2} \cite{Szafron-Collins}: A
Szekeres-Szafron metric is characterized by the following
conditions: (i) it is a perfect fluid solution with a geodesic and
hypersurface orthogonal unit velocity; (ii) each space-like
hypersurface orthogonal to the unit velocity has an induced
conformally flat metric $\gamma$; (iii) the Ricci tensor of $\gamma$
has a double eigenvalue; (iv) the shear tensor of the fluid has a
double eigenvalue.

Now, statements (i) and (iv) are first-order differential conditions
for the Ricci tensor of the spacetime metric. Moreover, as the
metric $\gamma$ is algebraic in the Ricci tensor, (ii) and (iii) are
differential constraints for the Ricci tensor, of third and second
order, respectively. Consequently, characterization 2 imposes third
order differential conditions solely in terms of the Ricci tensor.

Finally, a third invariant characterization was obtained by Barnes
and Rowlingson \cite{Barnes-Row} as a subclass of the irrotational
perfect fluid solutions with a purely electric Weyl tensor.
\ \\[2mm]
{\bf Invariant characterization 3} \cite{Barnes-Row}: A
Szekeres-Szafron metric is characterized by the following
conditions:
\begin{itemize}
\item[(i)]
It is a perfect fluid solution with a geodesic and irrotational unit
velocity.
\item[(ii)]
The Weyl tensor is purely electric and Petrov-Bel type D.
\item[(iii)]
The shear tensor of the fluid has a double eigenvalue, and its
associated eigenplane is the space-like principal plane of the Weyl
tensor.
\end{itemize}

The above constraints are algebraic conditions for the Weyl tensor
and first-order differential conditions for the Ricci tensor.

Takeno \cite{takeno} referred to a characterization in ``ideal form"
when a spacetime is labeled by equations exclusively involving
explicit concomitants of the metric tensor. He partially performed
this type of analysis for the spherically symmetric spacetimes
\cite{takeno, takeno-2}, a result we attained in two recent papers
\cite{fs-SSST, fs-ssst-Ricci}. This kind of IDEAL characterization
has also been achieved for other geometrically significant families
of metrics and for physically relevant solutions of the Einstein
equations \cite{fsI, fsS, fms, fsWEM, fsKY, fsIa, fsD, fsWEM2, cfs,
fsIb, fs-RainichEM-D, fsEM-sym, fsKerr, fswarped, fs-typeD,
IDEALcanepa}. The use of the appellation IDEAL (as an acronym) seems
to be adequate because the conditions obtained are Intrinsic
(depending only of the metric tensor), Deductive (not involving
inductive or inferential methods or arguments), Explicit (expressing
the solution non implicitly) and ALgorithmic (giving the solution as
a flow chart with a finite number of steps). The IDEAL approach
improves the previously known invariant characterizations since it
can be achieved by using the current tensor calculus packages.

The IDEAL characterization of a metric is based on the Cartan
historic results \cite{cartan} and it can be useful, at least, in
three fields of   theoretical physics. First, it allows us to check
whether a new solution to the Einstein equations is in fact already
known. Consequently, it is also a method to study the metric
equivalence problem, which is an alternative to the usual
Cartan-Brans-Karlhede approach \cite{brans, karlhede}. Second, it is
of interest in obtaining a fully algorithmic characterization of the
initial data which correspond to a given solution. So, our IDEAL
approaches to the Schwarzschild and Kerr solutions have been the
starting point in several papers \cite{garcia-parrado-vk,
garcia-parrado, garcia3}. And third, it has been proposed as a
fundamental tool in epistemic relativity for making gravimetry by
using relativistic positioning systems and relativistic stereometric
systems \cite{coll}.

Section \ref{sec-ideal} is devoted to the first goal of this paper:
provide an IDEAL labeling for the Szekeres-Szafron solutions. In
spite of the wide range of invariant studies on these metrics their
explicit expressions are not yet known.  Our starting point is the
invariant characterization 3 by Barnes and Rowlingson
\cite{Barnes-Row}, and we must obtain the Ricci and Weyl
concomitants that provide explicit expressions for the conditions
(i), (ii) and (iii).

In his seminal paper, Rainich \cite{Rainich} stated the problem of
obtaining the necessary and sufficient conditions for a metric to be
a solution to the non-null Einstein-Maxwell equations, and he solved
this problem in an IDEAL way. A similar study for the thermodynamic
perfect fluid solutions was presented in Ref.
\cite{Coll-Ferrando-termo}. By extension, we refer to a Rainich-like
approach to a family of solutions when we give an IDEAL
characterization of this family in terms of concomitants of the
Ricci tensor (see for example Ref. \cite{fs-RainichEM-D}). Our
second goal in this paper is to build the Rainich-like approach to
the Szekeres-Szafron solutions, a task accomplished in Sec.
\ref{sec-Rainich}. This study can be useful in practical situations
if we want to test if a metric defines a Szekeres-Szafron model by
using exclusively the unit velocity $u$, the energy density $\rho$
and the pressure $p$. By using only hydrodynamic variables the
Rainich-like approach also becomes useful from a conceptual point of
view.

The Szekeres-Szafron metrics admit the canonical form
\cite{Krasinski, Krasinski-Plebanski, Szafron, Szekeres}:
\begin{equation} \label{SS-canonica}
d s^2 = - dt^2 + e^{2 \alpha} d z^2 + e^{2 \beta} (d x^2 + d y^2) \,
,
\end{equation}
where the functions $\alpha = \alpha(t,z,x,y)$ and $\beta =
\beta(t,z,x,y)$ are submitted to the field equations. These
equations can be partially integrated by considering two classes,
$\beta,_z \not= 0$ (class I) and $\beta,_z =0$ (class II)
\cite{Krasinski, Krasinski-Plebanski}. In Sec. \ref{sec-beta} we
show that these coordinate conditions admit an invariant statement
with a specific geometric interpretation, and we give them as
explicit conditions for the Riemann tensor.

The Friedmann-Lema\^itre-Robertson-Walker (FLRW) universes can be
obtained from the Szekeres-Szafron spacetimes by means of a limiting
procedure \cite{Krasinski, Krasinski-Plebanski}. An intermediate
family of Szekeres-Szafron metrics, containing the FLRW universes,
comprises the geodesic perfect fluid solutions admitting a
three-dimensional group of isometries $O_3$ on space-like
two-dimensional orbits $S_2$. In fact, the two-surfaces $t=$const
and $z=$const of the metrics (\ref{SS-canonica}) of the
Szekeres-Szafron type have constant curvature and, consequently,
they admit a $G_3$. In Sec. \ref{sec-G3} we study when this isometry
group acts on the full spacetime and we offer the invariant and
explicit conditions characterizing this family.

The invariant constraint distinguishing class I and class II
Szekeres-Szafron metrics is a first-order differential condition
(which is linear in the first derivatives) on the Riemann tensor.
The condition characterizing the family with a $G_3$ on $S_2$ has
similar properties. In Sec. \ref{sec-class} we explore other
possible first-order differential invariant conditions, and we
comment on the invariant classes they define. These classes are also
labeled by explicit expressions.

Some remarks are presented in Sec. \ref{sec-termo} on the
interpretation of the Szekeres-Szafron models as a fluid in local
thermal equilibrium. We also offer an intrinsic and explicit
condition characterizing the thermodynamic Szekeres-Szafron
spacetimes.

A discussion on the results is given in Sec. \ref{sec-dis}. We also
offer one of all possible flow charts that can be built from our
IDEAL characterizations, and which can easily be realized by using
the current tensor calculus packages.

In Appendix \ref{app-A} we give the expressions of the coordinate
functions $\alpha$ and $\beta$ for the two classes I and II of the
Szekeres-Szafron metrics, and also the expressions of some scalar
invariants: pressure, energy density, Weyl eigenvalue and expansion.
In Appendix \ref{app-B} we prove some lemmas.

In this paper we work on an oriented spacetime with a metric tensor
$g$ of signature $\{-,+,+,+\}$. The Weyl tensor $W$, the Ricci
tensor $R$ and the scalar curvature $r$ are defined as given in Ref.
\cite{kramer}. For the metric product of two vectors, we write
$(x,y) = g(x,y)$, and we put $x^2 = g(x,x)$. The symbols $\nabla$
and $\nabla \cdot$ denote, respectively, the covariant derivative
and the divergence operator. For a $(p+1)$-tensor $P$ and a
$(q+1)$-tensor $Q$, $P \cdot Q$ denotes the $(p+q)$-tensor $(P \cdot
Q)_{\bar{p} \bar{q}}  = P_{\bar{p} \alpha} Q^{\alpha}_{\ \bar{q}}$,
with $\bar{p}$ and $\bar{q}$ denoting multi-indices. For a
two-tensor $T$, $T^2= T \cdot T$, $T(x)_{\alpha} = T_{\alpha \beta}
x^{\beta}$ and $T(x,y) = T_{\alpha \beta} x^{\alpha} y^{ \beta}$.
For a vector field $x$ and a function $f$ we write $x(f) =
x^{\alpha}
\partial_{\alpha} f$.

%%%%%%%%%%%%%%%%%%%%%%

\section{\label{sec-ideal}IDEAL characterization of the Szekeres-Szafron family of metrics}

In order to obtain an IDEAL labeling of the Szekeres-Szafron
solutions we must write the invariant conditions of characterization
3 in terms of explicit concomitants of the Ricci and Weyl tensors.
Condition (i) consists of the algebraic constraints which guarantee
the perfect fluid nature of the energy content, and the first-order
differential ones which impose the geodesic and irrotational
character of the fluid velocity.

The conditions for the Ricci tensor for a perfect fluid source were
obtained years ago \cite{bcm, Coll-Ferrando-termo}. Here we use a
minor adaptation of a more recent version \cite{fs-ssst-Ricci}:

\begin{proposition} \label{propo-fluper}
A spacetime is a perfect fluid solution if, and only if, the Ricci
tensor $R$ satisfies
\begin{equation} \label{fluper-conditions-A}
\Gamma^2 = \Gamma   , \quad  \Gamma(x,x) < 0   , \quad s \not = 0 \,
,
\end{equation}
where $x$ is any time-like vector, and
\begin{eqnarray} \label{fluper-definitions-A}
\ \Gamma \equiv  \frac14  g - \frac{1}{s} N   , \quad   N \equiv R -
\frac{r}{4} g, \\
s \equiv - 2 \sqrt[3]{\frac{\tr N^3}{3}} , \qquad  r\equiv \tr
R . \label{fluper-definitions-B}
\end{eqnarray}
Moreover, $\Gamma$ is the projector on the unit velocity $u$,
$\Gamma = - u \otimes u$, and the total energy $\rho$ and the
pressure $p$ of the fluid are given by
\begin{equation} \label{fluper-hydro}
 \rho = \frac14 (3 s + r)   , \qquad p = \frac14 (s - r)    .
\end{equation}
\end{proposition}

Now we must impose that the unit velocity has zero acceleration and
rotation, $a = w = 0$, and we must write these conditions in terms
of the Ricci concomitant $\Gamma =-  u \otimes u$. A straightforward
calculation leads to the following.
\begin{lemma} \label{lemma-a-w}
A unit time-like vector $u$ is geodesic and irrotational if, and
only if, the projector $\Gamma = - u \otimes u$ satisfies
\begin{equation} \label{a-w}
{\cal A} = 0 \, , \qquad  {\cal A}_{\alpha \beta \mu} \equiv
\nabla_{[ \alpha} \Gamma_{\beta ] \lambda}\,  \Gamma^{\lambda}_ {\
\mu}  \, .
\end{equation}
\end{lemma}

Thus, we have explicit expressions for the invariant condition (i).
Before studying condition (ii) we write the other kinematic
coefficients of the velocity $u$ in terms of $\Gamma$. Again, a
direct calculation allows us to obtain the following:
\begin{lemma} \label{lemma-theta-sigma}
If $u$ is a geodesic and irrotational unit time-like vector ($a =w =
0$) then
\begin{eqnarray} \label{theta}
 \theta u = \Theta \equiv      - \nabla \cdot \Gamma  , \qquad \\[1mm]
\sigma \otimes u =\Sigma    \equiv [\frac13  \gamma \otimes (\nabla
\cdot \Gamma) - \nabla \Gamma ] \cdot \Gamma , \ \ \gamma \equiv g
- \Gamma  . \qquad  \label{sigma}
\end{eqnarray}
\end{lemma}

Condition (ii) in the invariant characterization 3 imposes algebraic
constraints on the Weyl tensor. Nevertheless, if we calculate the
electric and magnetic parts with respect to an observer $u$ comoving
with the fluid, we obtain simpler explicit conditions involving both
the Weyl tensor and the algebraic Ricci concomitant $\Gamma$.
Indeed, remembering that a vanishing magnetic Weyl tensor implies
Petrov-Bel types I, D or O, then, a simple scalar condition
distinguishes the type D case. If we also take into account the
definition of the electric and magnetic Weyl tensors, we obtain the
following
\begin{lemma} \label{lemma-H}
Let $W$ be the Weyl tensor and $\Gamma = -u \otimes u$ the projector
on the unit time-like vector $u$. Then:
\begin{itemize}
\item[(i)]
The Weyl tensor $W$ is purely electric with respect the observer $u$
if, and only if,
\begin{equation} \label{H=0}
H = 0,   \qquad  H_{\alpha \beta} \equiv  \Gamma^{\lambda \mu}
(*W)_{\alpha \lambda \mu \beta}  \, .
\end{equation}
\item[(ii)]
The electric part $E$ of the Weyl tensor with respect to $u$ is
\begin{equation} \label{E}
  E_{\alpha \beta} \equiv  \Gamma^{\lambda \mu} W_{\alpha \lambda \mu \beta}  \, .
\end{equation}
\item[(iii)]
If $H=0$, the Weyl tensor is of type D or O if, and only if,
\begin{equation} \label{D}
I^3 = 6 J^2   ,   \quad \  I \equiv \tr E^2  , \quad  \ J  \equiv
\tr E^3    .
\end{equation}
The conformally flat case (type O) occurs if, and only if, $E=0$.
\end{itemize}
\end{lemma}

Condition (iii) in the invariant characterization 3 implies that the
shear $\sigma$ and the electric Weyl tensor $E$ have a common
eigenplane. But $E$ and $\sigma$ are traceless tensors and thus they
are, necessarily, proportional. Conversely, this last condition
implies that $\sigma$ admits an eigenplane provided that $E$ also
admits this eigenplane. Moreover, we can use the Ricci concomitant
$\Sigma$ defined in Eq. (\ref{sigma}) to write this condition and we
obtain the following
\begin{lemma} \label{lemma-sigma}
If $E$ admits a double eigenvalue, then $\sigma$ admits the same
eigenplane if, and only if,
\begin{equation} \label{E-sigma}
E_{\alpha \beta}\,  \Sigma_{\lambda \mu \nu} =  E_{\lambda \mu}\,
\Sigma_{\alpha \beta \nu}  .
\end{equation}
where $\Sigma$ is given in Eq. {\em (\ref{sigma})}.
\end{lemma}

Proposition \ref{propo-fluper} and lemmas \ref{lemma-a-w},
\ref{lemma-theta-sigma}, \ref{lemma-H} and \ref{lemma-sigma} make
explicit the three conditions in the invariant characterization 3.
Thus, we have performed the IDEAL labeling of the Szekeres-Szafron
solutions that we summarize in the following

\begin{theorem} \label{theorem-ideal}
A metric tensor $g$ is a Szekeres-Szafron solution if, and only if,
its Ricci tensor $R$ and Weyl tensor $W$ satisfy conditions {\em
(\ref{fluper-conditions-A}), (\ref{a-w}), (\ref{H=0}), (\ref{D})}
and {\em (\ref{E-sigma})}, where $\Gamma$ is given in Eq. {\em
(\ref{fluper-definitions-A})}, $s$ in Eq. {\em
(\ref{fluper-definitions-B})}, $E$ in Eq. {\em (\ref{E})}, and
$\Sigma$ in Eq. {\em (\ref{sigma})}.
\end{theorem}

Included in the Szekeres-Szafron family of metrics characterized
above are  two notable limits:  the nonexpanding solutions, which do
not properly correspond to a cosmological model, and the FLRW
universes, which occur in the conformally flat case or,
equivalently, in the shear-free limit. Lemmas
\ref{lemma-theta-sigma} and \ref{lemma-H} lead to the following

\begin{proposition}
Let $g$ be a Szekeres-Szafron solution characterized in theorem {\em
\ref{theorem-ideal}} and let $\Theta$ and $E$ and be the Riemann
concomitants defined in Eqs.{\em (\ref{theta})} and {\em (\ref{E})}.
Then:
\begin{itemize}
\item[(i)]
The metric $g$ defines a cosmological model with a nonvanishing
expansion if, and only if, $\Theta \not=0$.
\item[(ii)]
The metric $g$ defines a FLRW universe if, and only if, $\Theta \not
= 0$ and $E = 0$.
\end{itemize}
\end{proposition}

%%%%%%%%%%%%%%%%%%%%%%%%%%

\section{Characterization in terms of the Ricci tensor: Rainich-like approach}
\label{sec-Rainich}

In order to obtain an IDEAL labeling of the Szekeres-Szafron metrics
solely in terms of the Ricci tensor we could start from the
invariant characterization 2 which imposes third-order differential
conditions on the Ricci tensor, and then we should get explicit
expressions for them in order to perform a Rainich-like approach.
Nevertheless, we will prove here that starting from our IDEAL
labeling presented in the previous section and based on
characterization 3, we acquire a Rainich-like approach which imposes
second-order differential conditions on the Ricci tensor.

This approach is possible because the Ricci identities for the unit
velocity of the fluid give us the electric and magnetic Weyl tensors
in terms of second-order Ricci concomitants. Indeed, for a geodesic
an irrotational unit velocity, the Ricci identities imply
\cite{kramer}
\begin{eqnarray} \label{Ricci-E}
E = - \frac23 \theta \sigma - \dot{\sigma} - \sigma^2 + \frac13 \tr \sigma^2 \gamma  \, , \\[1mm]
H = {\rm curl} \, \sigma \, . \label{Ricci-H}
\end{eqnarray}
where a dot denotes the covariant directional derivative along the
unit velocity $u$.

Identity (\ref{Ricci-H}) shows that the algebraic invariant $H$
given in Eq. (\ref{H=0}), which depends on the Weyl and Ricci
tensors, is equal to a first-order differential invariant depending
solely on the Ricci tensor. We can write this Ricci invariant, which
we denote ${\cal H}$, in terms of $\Sigma$, and we obtain the
following
\begin{lemma} \label{lemma-Hcal}
Let $u$ be a geodesic and irrotational unit vector. The magnetic
part of the Weyl tensor with respect to $u$ vanishes if, and only
if,
\begin{equation} \label{calH=0}
{\cal H} = 0,   \qquad  {\cal H}^{\alpha \beta} \equiv
\eta^{\lambda \mu \nu (\alpha} \nabla_{\lambda} \Sigma^{\beta)}_{ \
\mu \nu}  \, ,
\end{equation}
where $\Sigma$ is given in Eq. {\em (\ref{sigma})} and $\Gamma= - u
\otimes u$.
\end{lemma}

Similarly, identity (\ref{Ricci-E}) shows that the invariant $E$
given in Eq. (\ref{E}), which depends algebraically on the Weyl and
Ricci tensors, is equal to a second-order differential invariant
depending solely on the Ricci tensor. We could write this Ricci
invariant in terms of $\Theta$ and $\Sigma$, and impose on it the
constraints (\ref{D}) and (\ref{E-sigma}). This procedure leads to
tangled conditions with nonlinear second-order terms. Nevertheless,
we can obtain simpler conditions by imposing on the shear $\sigma$
the condition of having a double eigenvalue. A straightforward
calculation allows us to write this condition in terms of $\Sigma$
and we get the following
\begin{lemma} \label{lemma-sigmaD}
A geodesic and irrotational unit vector $u$ has a shear with a
double eigenvalue if, and only if,
\begin{equation} \label{sigma-D}
6 \, {\cal S}^2 = S^3  ,  \ \ {\cal S}_{\alpha} \equiv
\Sigma_{\lambda  \beta \gamma}  \Sigma_{\mu}^{\ \beta \gamma}
\Sigma^{\lambda \mu} _{\ \ \alpha}  ,  \  \ S \equiv
\Sigma_{\lambda \mu \nu}  \Sigma^{\lambda \mu \nu} \! . \
\end{equation}
where $\Sigma$ is given in Eq. {\em (\ref{sigma})} and $\Gamma= - u
\otimes u$.
\end{lemma}

On the other hand, we must impose the proportionality of $\sigma$
and $E$. From Eq. (\ref{Ricci-E}) this condition is equivalent to
the proportionality of $\sigma$ and $\dot{\sigma}$. If we write this
condition in terms of $\Sigma$ we obtain the following
\begin{lemma} \label{lemma-sigma-punt}
For a geodesic and irrotational unit vector $u$, the shear $\sigma$
and its  derivative $\dot{\sigma}$ are proportional tensors if, and
only if,
\begin{equation} \label{sigma-punt}
\nabla_{\nu} \Sigma_{\alpha \beta}^{\ \ \nu}\, \Sigma_{\lambda \mu
\gamma} = \nabla_{\nu} \Sigma_{\lambda \mu}^{\ \ \nu}\,
\Sigma_{\alpha \beta \gamma} \, .
\end{equation}
where $\Sigma$ is given in Eq. {\em (\ref{sigma})} and $\Gamma= - u
\otimes u$.
\end{lemma}

Proposition \ref{propo-fluper} and lemmas \ref{lemma-Hcal},
\ref{lemma-sigmaD} and \ref{lemma-sigma-punt} allow us to replace
conditions (\ref{H=0}), (\ref{D}) and (\ref{E-sigma}) in theorem
\ref{theorem-ideal} with conditions (\ref{calH=0}), (\ref{sigma-D})
and (\ref{sigma-punt}). This way, the first-order conditions on the
Weyl and Ricci tensors can be replaced by second-order ones
involving only the Ricci tensor. Consequently, we   acquire  the
following Rainich-like characterization:

\begin{theorem} \label{theorem-Rainich}
A metric tensor $g$ is a Szekeres-Szafron solution if, and only if,
its Ricci tensor $R$ satisfies conditions {\em
(\ref{fluper-conditions-A}), (\ref{a-w}), (\ref{calH=0}),
(\ref{sigma-D})} and {\em (\ref{sigma-punt})}, where $\Gamma$ is
given in Eq. {\em (\ref{fluper-definitions-A})}, $s$ in Eq. {\em
(\ref{fluper-definitions-B})}, and $\Sigma$ in Eq. {\em
(\ref{sigma})}.
\end{theorem}

The FLRW universes limit also admits a characterization in terms of
the Ricci tensor:

\begin{proposition}
A Szekeres-Szafron solution characterized in theorem {\em
\ref{theorem-Rainich}} becomes  a FLRW universe if, and only if,
$\Theta \not = 0$ and $\Sigma = 0$, where $\Theta$ and $\Sigma$ are
given in Eqs. {\em (\ref{theta})} and {\em (\ref{sigma})}.
\end{proposition}

%%%%%%%%%%%%%%%%%%%%%%%%%

\section{\label{sec-beta}IDEAL labeling of both families of Szekeres-Szafron metrics}
Szafron and Collins \cite{Szafron-Collins} showed that the
coordinate condition $\beta,_z =0$, which defines the class II of
the Szekeres-Szafron metrics, admits an invariant statement for the
{\em strict} Szekeres-Szafron (sSS) metrics, that is, when $\sigma
\not=0$ (the spacetime is not  a FLRW  universe). Here, to provide
an interpretation of this invariant condition and to obtain an
explicit expression of it in terms of the curvature tensor, we
analyze the geometric properties of the unitary eigenvector $b$
associated with the simple eigenvalue of the electric Weyl tensor.

In the canonical coordinate system (\ref{SS-canonica}) $b$ and its
covariant derivative take the forms
\begin{eqnarray} \label{v-nabla-v}
b = e^{\alpha} d z  , \ \  \nabla b =  b \otimes a_b + \frac12 \theta_b  h  , \  \ h \equiv \gamma - b \otimes b  , \quad \  \\[1mm]
a_b \equiv \nabla_b b = \dot{\alpha} u - h(d \alpha)  ,  \  \ \theta_b
\equiv \nabla \cdot b = 2 \, e^{-\alpha} \beta,_z .   \quad \
\label{a-theta-v}
\end{eqnarray}
For a function $f$, $\dot{f} = u(f) = u^{\alpha} \partial_{\alpha}f
= f,_t$. Note that $h$ is the projector on the space-like principal
plane of the Weyl tensor, and $a_v$ and $\theta_v$ are,
respectively, the acceleration and the expansion of the eigenvector
$b$. Equation (\ref{v-nabla-v}) shows that $b$ is expansion-free if,
and only if, it is shear-free. Then, Eq. (\ref{a-theta-v}) implies
the following
\begin{lemma} \label{lemma-b'=0}
For the strict Szekeres-Szafron solutions we have the following three equivalent conditions:
\begin{itemize}
\item[(i)]
The metric is of class II ($\beta,_z = 0$).
\item[(ii)]
The simple eigenvector of the electric Weyl tensor is
expansion-free.
 \item[(ii)]
The simple eigenvector of the electric Weyl tensor is shear-free.
\end{itemize}
\end{lemma}

On the other hand, the projectors on the direction defined by the
simple eigenvector, $B = b \otimes b$, and on the space-like
principal plane, $h$, can be obtained as
\begin{equation} \label{B}
B \equiv \frac13 \left(\frac{2}{\omega} E + \gamma\right),
\quad \ h \equiv \gamma -  B  , \quad \ \omega \equiv \frac{2 J}{I}
\, ,
\end{equation}
where $\gamma$, $E$, $I$ and $J$ are given in Eqs. (\ref{sigma}),
(\ref{E}) and (\ref{D}). Note that $\omega$ is the simple Weyl
eigenvalue. Finally, if we write condition $\theta_b = 0$ in terms
of $B$, and we take into account lemma \ref{lemma-b'=0}, we obtain
the following
\begin{proposition} \label{prop-II}
Let $g$ be a Szekeres-Szafron solution characterized in theorem {\em
\ref{theorem-ideal}}. Then, if $E \not= 0$, $g$ is of class II if,
and only if, the Ricci and Weyl tensors satisfy
\begin{equation} \label{SS-II}
B(\nabla \cdot B) = 0 \, ,
\end{equation}
where $B$ is given in Eq. {\em (\ref{B})}.
\end{proposition}

In order to obtain a Rainich-like labeling of the Szekeres-Szafron
metrics of class I and class II, we must write the condition
$\theta_b=0$ only in terms of the Ricci tensor. We know that $b$ is
also the eigenvector of $\sigma$ associated with the simple
eigenvalue. Then, from the expressions (\ref{sigma}) and
(\ref{sigma-D}) we obtain
\begin{eqnarray} \label{calB}
b \otimes b \otimes u = {\cal B} \equiv -\frac{1}{\sqrt{{-\cal
S}^2}}[S \, \Sigma + \gamma \otimes {\cal S}] ,
\\
h_{\alpha \beta} = \gamma_{\alpha \beta} + {\cal B}_{\alpha}^{\
\lambda \mu} {\cal B}_{\beta \lambda \mu} , \label{hache}
\end{eqnarray}
where $\Sigma$, $\gamma$, ${\cal S}$ and $S$ are given in Eqs.
(\ref{sigma}) and (\ref{sigma-D}). Finally, if we write the
condition $\theta_b = 0$ in terms of ${\cal B}$, and we take into
account lemma \ref{lemma-b'=0}, we obtain the following
\begin{proposition} \label{prop-II-shear}
Let $g$ be a Szekeres-Szafron solution characterized in theorem {\em
\ref{theorem-Rainich}}. Then, if $\Sigma \not= 0$, $g$ is of class
II if, and only if, the Ricci tensor satisfies
\begin{equation}
{\cal B}_{\alpha \lambda \mu} \, (\nabla \cdot {\cal B})^{\lambda
\mu} = 0 \, ,
\end{equation}
where ${\cal B}$ is given in Eq. {\em (\ref{calB})}.
\end{proposition}
%

 %%%%%%%%%%%%%%%%%%%%%

\section{Invariant and IDEAL labeling of the Szekeres-Szafron metrics admitting a $G_3$ on $S_2$.}
\label{sec-G3}

It is known \cite{Krasinski-Plebanski} that the two-surfaces
$t=$const, $z=$const of the Szekeres-Szafron spacetimes have
constant curvature, that is, they admit a maximal group of
isometries $G_3$. From the coordinate expressions given in Appendix
\ref{app-A} we can analyze when this group is a group of isometries
of the full spacetime. We have the following result:
\begin{lemma} \label{lemma-G3}
For a strict Szekeres-Szafron solution the three
following conditions are equivalent:
\begin{itemize}
\item[(i)]
It admits a three-dimensional group $G_3$ on two-dimensional
space-like orbits $S_2$.
\item[(ii)]
$\alpha,_x = \alpha,_y = 0$ ($h(d \alpha) =0$).
\item[(iii)]
$\nu,_{zx} = \nu,_{zy} = 0$ for class I ($\beta,_z \not= 0$), or
$P,_{x} = P,_{y} = 0$ for class II ($\beta,_z = 0$).
\end{itemize}
\end{lemma}
From Eq. (\ref{SS-canonica}) and expressions in Appendix
\ref{app-A}, it is evident that (i) implies (ii), and (ii) implies
(iii). It is also trivial that (iii) leads to (i) for the class-II
Szekeres-Szafron metrics. For class I, condition (iii) implies $(\ln
{\rm S}),_{xz} = (\ln {\rm S}),_{yz} = 0$, and then the functions
$U(z)$, $V_1(z)$, $V_2(z)$ and $W(z)$ in the expression (\ref{I-S})
of ${\rm S}(z,x,y)$ differ   by a constant. Then, a linear change in
the $(x,y)$ coordinates leads to ${\rm S}(z,x,y)= U(z) C(x,y)$,
where $C(x,y)$ takes the form (\ref{II-C-S}), and then (i) follows.

The invariant condition $h(a_b)=0$ is equivalent to $h(d \alpha)=0$
as a consequence of Eq. (\ref{a-theta-v}). Thus the lemma above
implies that this invariant condition characterizes the sSS metrics
admitting a three-dimensional group of isometries $G_3$ with
space-like two-dimensional orbits $S_2$. Furthermore, the null Weyl
principal directions are $\ell_{\pm} = u \pm b = - dt \pm \alpha d
z$, and a straightforward calculation shows that they are tangent to
null geodesics if, and only if, $h(d \alpha)=0$. Moreover, condition
$h(a_b)=0$ can be written in terms of one of the Riemann
concomitants $B$ and ${\cal B}$ defined in Eqs. (\ref{B}) and
(\ref{calB}), respectively.  Consequently, we have the following
\begin{proposition} \label{propo-G3}
For a strict Szekeres-Szafron metric the following conditions are
equivalent:
\begin{itemize}
\item[(i)]
It admits a three-dimensional group $G_3$ on two-dimensional
space-like orbits $S_2$.
\item[(ii)]
The null Weyl principal directions define geodesic congruences.
\item[(iii)]
$h(a_b)=0$, where $a_b$ is the acceleration of the simple Weyl
eigenvector $b$.
\item[(iv)]
$h(\nabla \cdot B) = 0$, where $h$ and $B$ are given in Eq. {\em
(\ref{B})}.
\item[(v)]
$h \cdot (\nabla \cdot {\cal B}) = 0$, where ${\cal B}$ and $h$ are
given in   Eqs. {\em (\ref{calB})} and {\em (\ref{hache})}
respectively.
\end{itemize}
\end{proposition}

Condition (ii) was introduced by Wainwright \cite{Wainwright-b} in
classifying the type D perfect fluid solutions, showing that if one
of the two null principal directions is geodesic then both are.
Moreover, he proved \cite{Wainwright} that  (ii) is a sufficient
condition for the existence of a $G_3$ on $S_2$ in the case of a
dust Szekeres metric.

Our proposition \ref{propo-G3} states that this condition is also a
necessary one, and we extend the result to the full set of the sSS
metrics. Conditions (ii) and (iii) are two equivalent invariant
conditions which characterize the Szekeres-Szafron metrics admitting
a $G_3$ on $S_2$. Furthermore, condition (iv) is explicit in both
the Weyl and the Ricci tensor, and (v) is explicit in the Ricci
tensor.

Proposition 6 implies that the sSS metrics with $h(a_b) = 0$ are
perfect fluid solutions with geodesic velocity admitting a $G_3$ on
$S_2$. Conversely, the existence of this group of isometries implies
that the spacetime is Petrov-Bel type D, the space-like principal
plane being tangent to the group orbits and an eigenplane of the
shear. Also the fluid is irrotational. Consequently, under the
geodesic constraint, all the conditions in the invariant
characterization 3 of the Szekeres-Szafron metrics hold. Thus, we
have the following
\begin{proposition} \label{propo-G3-geodesic}
The Szekeres-Szafron metrics with $h(a_b) = 0$ are the perfect fluid
solutions with geodesic velocity admitting a group of isometries
$G_3$ on orbits $S_2$.
\end{proposition}

In the next section we will prove that there are other invariant
conditions, which are also linear in the first derivatives of the
Riemann tensor, that either imply or are equivalent to the existence
of a $G_3$ on $S_2$. Now we analyze the second-order condition $h(d
\theta)=0$. From  Eq. (\ref{expansion-I}) (respectively, Eq.
(\ref{expansion-II})) of the expansion of the fluid of a sSS metrics
of class I (respectively, class II), we obtain that $h(d \theta) =
0$ implies $(\dot{\phi}\phi_z - \phi \dot{\phi},_z) h(d \nu_z)=0$
(respectively, $(\dot{\phi}\lambda - \phi \dot{\phi\lambda},_z) h(d
P)=0$). The factor $\dot{\phi}\phi_z - \phi \dot{\phi},_z$
(respectively, $\dot{\phi}\lambda - \phi \dot{\phi\lambda},_z$) only
vanishes in the FLRW limit. Thus, for the sSS metrics $h(d \theta) =
0$ implies that $\nu,_{zx} = \nu,_{zy} = 0$ (respectively,
$P_x=P_y=0$) and, as a consequence of lemma \ref{lemma-G3}, the
spacetime admits a $G_3$ on $O_2$. Conversely, this condition
implies $h(d \theta) = 0$ since any Riemann scalar is invariant by
the group. Thus, we can state the following

\begin{proposition} \label{propo-h-theta}
A strict Szekeres-Szafron metric admits an isometry group $G_3$ on
orbits
$O_2$ if, and only if, the invariant condition $h(d \theta) = 0$ holds.\\
The explicit expression for this constraint is $h \cdot \nabla
\Theta \cdot \Gamma =0$, where $\Gamma$ is given in Eq. {\em
(\ref{fluper-definitions-A})}, $\Theta$ in Eq. {\em (\ref{theta})},
and $h$ in Eq. {\em (\ref{B})}.
\end{proposition}
As a consequence of the proposition above, the Szekeres-Szafron
solutions with homogeneous expansion, $d \theta \wedge u = 0$, admit
a $G_3$ on $S_2$.

 %%%%%%%%%%%%%%%%%%%%%

\section{\label{sec-class}Other Szekeres-Szafron subclasses defined by first-order differential invariant conditions}

The invariant tensor $B$ defined in Eq. (\ref{B}) is an algebraic
Riemann invariant. Consequently, constraint (\ref{SS-II}) that
characterizes the sSS metrics of class II is a first-order
differential condition, which is linear in the first derivatives.
The constraint $h(\nabla \cdot B) = 0$ that characterizes the sSS
admitting a $G_3$ on $S_2$ has similar qualities. Any Riemann
invariant with analogous differential properties will generate a
classification of these metrics. In order to index all these
invariants let us notice first the algebraic Riemann invariants:
{\em unit velocity $u$, energy density $\rho$, pressure $p$, simple
Weyl eigenvalue $\omega$, and simple associated eigenvector $b$}.

Then, the first-order differential invariants (which are linear in
the first derivatives) are defined by the kinematic coefficients of
the invariant vectors $u$ and $b$, and the projection on $u$, $b$
and $h$ (projector on the space-like eigenplane) of the gradient of
the scalar invariants $\rho$, $p$ and $\omega$. For the
Szekeres-Szafron metrics some of these invariants identically
vanish. Moreover, $\theta_b=0$ if, and only if, $\sigma_b =0$ (lemma
\ref{lemma-b'=0}), and then it is enough to consider any of these
two invariants. Furthermore, the acceleration $a_b$ can be projected
on $u$ and on the space-like Weyl principal plane. Now we analyze
the classification induced by each of these differential invariants.
%

%%%%%%%%%%%%%%%%%
\subsection{Classes defined by invariants associated with the unit velocity $u$}

The only nonvanishing coefficients are the shear $\sigma$ and the
expansion $\theta$. We know that the first-order condition
$\sigma=0$ is equivalent to the algebraic one $E=0$, and it leads to
the FLRW universes. In this case the algebraic invariants $b$ and
$\omega$ are not defined. Otherwise, when $\sigma \not=0$, we have
the sSS metrics, a set where the other first-order invariants induce
a classification that we consider in the following subsections.

Moreover, as a consequence of the conservation equation, $\dot{\rho}
+ (\rho + p) \theta = 0$, $\theta =0$ is a constraint equivalent to
$u(d \rho) = \dot{\rho} =0$, a condition that we analyze below for
the sSS metrics.

Note that both conditions, $\sigma = 0$ and $\theta=0$ admit an
explicit expression in terms of the Ricci invariants  defined in
Eqs. (\ref{sigma}) and (\ref{theta}): $\Sigma=0$ and $\Theta=0$.
%

%%%%%%%%%%%%%%%%%

\subsection{Classes defined by invariants associated with the simple eigenvector $b$}

Now we can consider the expansion $\theta_b$ and the projections
$(a_b, u)$, $h(a_b)$ of the acceleration $a_b$ of the vector $b$.

In lemma \ref{lemma-b'=0} of Sec. \ref{sec-beta} we have shown that
the condition $\theta_b=0$ characterizes the sSS metrics of class
II, and we have also presented ideal statements for this condition
(propositions \ref{prop-II} and \ref{prop-II-shear}).

Furthermore, the analysis presented in the previous section shows
that the condition $h(a_b)=0$ characterizes the sSS metrics
admitting an isometry group $G_3$ on orbits $S_2$, and we have
presented ideal statements for this condition (proposition
\ref{propo-G3}).

Finally, the invariant condition $(a_b, u)=0$ is equivalent to
$\alpha,_t = 0$ as a consequence of Eq. (\ref{a-theta-v}). For  sSS
metrics of class I (respectively, class II), Eqs.
(\ref{I-alpha-beta}) and (\ref{I-S}) (respectively, Eqs.
(\ref{II-alpha-beta}) and (\ref{II-phi-S})) imply that this
condition leads to $\dot{\phi},_{z} + \dot{\phi} \nu,_z=0$
(respectively, $\dot{\lambda} + P \dot{\phi} =0$) and then
$\dot{\phi} \nu,_{zx}= \dot{\phi} \nu,_{zy}=0$ (respectively,
$\dot{\phi} P,_x = \dot{\phi} P,_y=0$). Thus, either $\dot{\phi} =
0$ (respectively, $\dot{\phi} = \dot{\lambda} =0$) and the metric is
static, or $\nu,_{zx} = \nu,_{zy} = 0$ (respectively, $P,_x =
P,_y=0$) and the spacetime admits a $G_3$ on $S_2$ (see lemma
\ref{lemma-G3}). Note that the static condition is not possible in
the sSS metrics. On the other hand, we can write the invariant
condition $(a_b, u)=0$ in terms of the explicit Riemann concomitants
$B$ and ${\cal B}$. Thus, we obtain the following
\begin{proposition} \label{propo-abpunt}
For a strict Szekeres-Szafron metric the following conditions are
equivalent:
\begin{itemize}
\item[(i)] $(a_b,u)=0$.
\item[(ii)]
$\Gamma(\nabla \cdot B) = 0$ with  $B$     given in Eq. {\em
(\ref{B})}, and $\Gamma$  in Eq. {\em (\ref{fluper-definitions-A})}.
\item[(iii)]
$\Gamma \cdot (\nabla \cdot {\cal B}) = 0$ with   ${\cal B}$ given
in Eq. {\em (\ref{calB})} and $\Gamma$ in Eq. {\em
(\ref{fluper-definitions-A})}.
\end{itemize}
Moreover, any of these conditions implies $h(a_b)=0$, that is, the
spacetime admits necessarily an isometry group $G_3$ on orbits
$S_2$.
\end{proposition}
Note that condition (ii) is an explicit expression of (i) in terms
of the Ricci and Weyl tensors, and (iii) is its explicit expression
only in terms of the Ricci tensor.

\subsection{Classes defined by invariants associated with the pressure $p$}

For the sSS metrics we have $dp \wedge u = 0$, and then the sole
scalar $u( p) = \dot{p}$ can be considered. If it vanishes, $u (p
)=0$, we have a constant pressure. This class, which includes the
Szekeres dust solutions, has been  widely studied by several authors
(see Refs. \cite{Krasinski, Krasinski-Plebanski, hellaby,G-hellaby}
and references therein).

As a consequence of Eq. (\ref{fluper-hydro}), condition $u (p )=0$
admits an explicit expression in terms of Ricci invariants: $d r = d
s$, where $r$ and $s$ are given in Eq. (\ref{fluper-definitions-B}).

%%%%%%%%%%%%%%%%%

\subsection{Classes defined by invariants associated with the simple Weyl eigenvalue $\omega$}

Now we have the invariants defined by the projections $u(\omega)$,
$b(\omega)$, $h(d \omega)$ of the gradient $d \omega$ of the scalar
$\omega$.

First we analyze the invariant condition $h(d \omega) = 0$. From Eq.
(\ref{eigenvalue-I}) (respectively, Eq. (\ref{eigenvalue-II})) of
the Weyl eigenvalue of a sSS metrics of class I (respectively, class
II) we obtain $h(d \omega) = - \omega \frac{\phi}{\phi_z + \phi
\nu_z} h(d \nu_z)$ (respectively, $h(d \omega) = - \omega
\frac{\phi}{\lambda + \phi P} h(d P))$. Thus $h(d \omega) = 0$
implies that $\nu,_{zx} = \nu,_{zy} = 0$ (respectively, $P_x=P_y=0$)
and, as a consequence of lemma \ref{lemma-G3}, the spacetime admits
a $G_3$ on $O_2$. Conversely, this condition implies $h(d \omega) =
0$ since any Riemann scalar is invariant by the group. Thus, we can
state the following

\begin{proposition} \label{propo-h-omega}
A strict Szekeres-Szafron metric admits an isometry group $G_3$ on
orbits
$O_2$ if, and only if, the invariant condition $h(d \omega) = 0$ holds.\\
The explicit expressions of the involved Riemann invariants $h$ and
$\omega$ are given in Eq. {\em (\ref{B})}.
\end{proposition}

Second  we analyze the invariant condition $u(\omega) = 0$, that is,
$\dot{\omega}=0$. From Eq.  (\ref{eigenvalue-I}) (respectively, Eq.
(\ref{eigenvalue-II})) of the Weyl eigenvalue of a sSS metric of
class I (respectively, class II) we obtain $h(d \dot{\omega}) = -
\omega \,
\partial_t\! \left(\frac{\phi}{\phi_z + \phi \nu_z}\right) h(d
\nu_z)$ (respectively, $h(d \dot{\omega}) = - \omega \, \partial_t\!
\left(\frac{\phi}{\lambda + \phi P}\right) h(d P)$). The factor
$\partial_t\! \left(\frac{\phi}{\phi_z + \phi \nu_z}\right)$
(respectively, $\partial_t\! \left(\frac{\phi}{\lambda + \phi
P}\right)$) only vanishes in the FLRW limit. Thus, for the sSS
metrics $u(\omega) = 0$ implies that $\nu,_{zx} = \nu,_{zy} = 0$
(respectively, $P_x=P_y=0$) and, as a consequence of lemma
\ref{lemma-G3}, the spacetime admits a $G_3$ on $O_2$. Thus, we can
state the following
\begin{proposition} \label{propo-omegapunt}
For a strict Szekeres-Szafron metric the invariant condition
$u(\omega)=0$ implies $h(d \omega) = 0$, that is, the spacetime
admits an isometry group $G_3$ on orbits $O_2$. Moreover, the
condition $u(\omega)=0$ admits the explicit statement $\Gamma(d
\omega) =0$, where $\Gamma$ is given in Eq. {\em
(\ref{fluper-definitions-A})} and $\omega$ in Eq. {\em (\ref{B})}.
\end{proposition}

Now we need to analyze the invariant condition $b(\omega) = 0$, that
is, $\omega,_z = 0$. For a sSS metric of class I, lemma
\ref{lemma-omega-r-I} in Appendix \ref{app-B} implies that
$\nu,_{zx} = \nu,_{zy}  = 0$ and, as a consequence of lemma
\ref{lemma-G3}, the spacetime admits a $G_3$ on $O_2$. For a sSS
metric of class II, lemma \ref{lemma-omega-r-II} in Appendix
\ref{app-B} implies that $\xi =
\partial_z$ is a Killing vector. Thus, we have the following
\begin{proposition} \label{propo-b-omega}
For a strict class-I Szekeres-Szafron metric the invariant condition $b(\omega)=0$ implies $h(d \omega) = 0$,
that is, the spacetime admits an isometry group $G_3$ on orbits $O_2$.\\
For a strict class-II Szekeres-Szafron metric the invariant condition $b(\omega)=0$ implies that the Weyl eigenvector
 $b$ determines a Killing vector direction. \\
Moreover, the condition $b(\omega)=0$ admits the explicit statement
$B(d \omega) =0$, where $B$ and $\omega$ are given in Eq. {\em
(\ref{B})}.
\end{proposition}
%

%%%%%%%%%%%%%%%%%%%%%

\subsection{Classes defined by invariants associated with the energy density $\rho$}

Now we have the projections $u(\rho)$, $b(\rho)$ and $h(d \rho)$ of
the gradient $d \rho$ of the the scalar $\rho$. From the expressions
of the pressure $p$, the energy density $\rho$ and the Weyl
eigenvalue $\omega$ given in Appendix \ref{app-A} we obtain
\begin{equation} \label{rho-omega}
\frac13 \rho + p + 2  \omega = - 2 \frac{\ddot{\phi}}{\phi} \, .
\end{equation}
From this relation we have that $h(d \rho)=0$ if, and only if, $h(d
\omega)=0$. Consequently, proposition \ref{propo-h-omega} applies
and we can state the following
\begin{proposition} \label{propo-h-rho}
A strict Szekeres-Szafron metric admits an isometry group $G_3$ on
orbits
$O_2$ if, and only if, the invariant condition $h(d \rho) = 0$ holds.\\
The explicit expressions of the involved Riemann invariants $h$ and
$\rho$ are given in Eqs.  {\em (\ref{B})} and {\em
(\ref{fluper-hydro})}, respectively.
\end{proposition}

Now we analyze the invariant condition $u(\rho) = 0$, that is,
$\dot{\rho}=0$ or, equivalently $\theta=0$. From Eq.
(\ref{expansion-I}) (respectively, Eq. (\ref{expansion-II})) of the
expansion of a sSS metrics of class I (respectively, class II) we
obtain $\phi \dot{\phi},_r + 3 \phi \dot{\phi} \nu,_r + 2 \dot{\phi}
\phi,_r =0$ (respectively, $\phi \dot{\lambda} + 3 \phi \dot{\phi} P
+ 2 \dot{\phi} \lambda =0$) and, differentiating with respect to $x$
and $y$, we have that $\nu,_{zx} = \nu,_{zy} = 0$ (respectively,
$P_x=P_y=0$) and, as a consequence of lemma \ref{lemma-G3}, the
spacetime admits a $G_3$ on $O_2$. Thus, we can state the following
\begin{proposition} \label{propo-rhopunt}
For a strict Szekeres-Szafron metric the invariant condition
$u(\rho)=0$ implies $h(d \rho) = 0$, that is, the spacetime admits
an isometry group $G_3$ on orbits $O_2$. Moreover, the condition
$u(\rho)=0$ admits the explicit statement $\Gamma(d \rho) =0$, where
$\Gamma$ is given in Eq. {\em (\ref{fluper-definitions-A})} and
$\rho$ in Eq. {\em (\ref{fluper-hydro})}.
\end{proposition}

In the case of the invariant condition $b(\rho)=0$, that is $\rho,_z
=0$, we must distinguish between class I and class II metrics. In
the second case $\phi=\phi(t)$ and then Eq. (\ref{rho-omega})
implies that $b(\rho)=0$ if, and only if, $b(\omega)=0$, and then
proposition \ref{propo-b-omega} applies and the spacetime admits a
Killing vector which is parallel to $b$. In the first case, for
metrics of class I, the spacetime is a FLRW universe as a
consequence of lemma \ref{lemma-I-rhoz} in Appendix \ref{app-B}.
Thus, we have the following
\begin{proposition} \label{propo-b-rho}
Strict Szekeres-Szafron metrics of type I fulfilling the invariant condition $b(\rho)=0$ do not exist.\\
For a strict Szekeres-Szafron metric of type II the invariant condition $b(\rho)=0$ implies that the Weyl eigenvector $b$ determines a Killing vector direction. \\
Moreover, the condition $b(\rho)=0$ admits the explicit statement
$B(d \rho) =0$, where $B$ and $\rho$ are given in Eqs. {\em
(\ref{B})} and {\em (\ref{fluper-hydro})}, respectively.
\end{proposition}

%%%%%%%%%%%%%%%%

\section{\label{sec-termo}On the thermodynamic Szekeres-Szafron solutions}

A relevant step in studying perfect fluid solutions is to analyze
their interpretation as reasonable physical media. The Pleba\'nski
\cite{Plebanski} energy conditions are necessary algebraic
conditions for physical reality and, in the perfect fluid case, they
state  $-\rho < p \leq \rho$. The determination of the spacetime
regions where these constraints hold is a basic query in analyzing a
given perfect fluid solution.

Furthermore, if we want the solution to describe a perfect fluid in
local thermal equilibrium we must impose complementary constraints.
A necessary condition for the fluid to admit a thermodynamic scheme
is that a function $n$ exists such that \cite{fluperLTE}
\begin{equation} \label{lte-n}
\dot{n} + n \theta =0 \, , \qquad   dn \wedge d p \wedge d \rho = 0
\, .
\end{equation}
Then, the function of state $n=n(\rho,p)$ is the conserved matter
density of the fluid.

It is worth remarking that Eq. (\ref{lte-n}) is not an {\em
intrinsic} condition on a perfect energy tensor $T \equiv (u, \rho,
p)$ in order to represent the energetic evolution of a perfect fluid
in local thermal equilibrium. Indeed, it involves the function $n$,
which is not defined by $T$. In Ref. \cite{Coll-Ferrando-termo} (see
also Ref. \cite{fluperLTE}) we presented an intrinsic and explicit
condition: {\em a nonisoenergetic ($\dot{\rho} \not= 0$) perfect
energy tensor $T$ evolves in local thermal equilibrium if, and only
if, the hydrodynamic variables $(u, \rho, p)$ fulfill}
\begin{equation} \label{lte-chi}
\   d \chi \wedge d p \wedge d \rho = 0 \, , \qquad \chi \equiv
\frac{\dot{p}}{\dot{\rho}}   \, .
\end{equation}
Then the {\em indicatrix of the local thermal equilibrium} $\chi$ is
a function of state, $\chi = \chi(\rho,p)$, which represents the
square of the speed of sound \cite{fluperLTE}. This IDEAL
characterization of local thermal equilibrium enabled us to
construct a Rainich-like theory for the thermodynamic perfect fluids
\cite{Coll-Ferrando-termo}. To get this, it is enough to write
condition (\ref{lte-chi}) in terms of the Ricci tensor and to add it
to the conditions in proposition \ref{propo-fluper}
\cite{Coll-Ferrando-termo} \cite{fs-ssst-Ricci}.

In the case of the Szekeres-Szafron spacetimes this thermodynamic
condition admits an equivalent and simpler expression. Indeed as $p$
is a function of $t$, we have $d p \wedge u = d \dot{p} \wedge u
=0$. Then, Eq. (\ref{lte-chi}) is equivalent to $d \dot{\rho} \wedge
d \rho \wedge u = 0$. Also $\dot{\rho}$ can be substituted by
$\theta$ because of the conservation equation $\dot{\rho} + (\rho +
p) \theta = 0$. Thus, Eq. (\ref{lte-chi}) becomes
\begin{equation} \label{lte-theta}
\   d \theta \wedge d \rho \wedge u = 0 \,  .
\end{equation}
Moreover, we can substitute $\rho$ by any of the Ricci scalar
invariants, $r$ and $s$, defined in Eq.
(\ref{fluper-definitions-B}). The resulting constraint can be
written in terms of $\Gamma$ and $\Theta$, and we obtain the
following
\begin{proposition} \label{prop-termo}
Let $g$ be a Szekeres-Szafron solution characterized in theorem {\em
\ref{theorem-ideal}} or in theorem \ref{theorem-Rainich}. Then, the
solution represents a perfect fluid in local thermal equilibrium if,
and only if, the Ricci tensor satisfies
\begin{equation} \label{SS-termo}
{\cal T} = 0 \, , \qquad  {\cal T}^{\alpha} \equiv \eta^{\alpha
\lambda \mu \nu} \Gamma_{\lambda}^{\  \beta} \partial_{\mu} r
\nabla_{\nu}  \Theta_{\beta}   \, .
\end{equation}
where $\Gamma$ is given in Eq. {\em (\ref{fluper-definitions-A})},
$r$ in Eq. {\em (\ref{fluper-definitions-B})} and $\Theta$ in Eq.
{\em (\ref{theta})}.
\end{proposition}

The above IDEAL labeling of the thermodynamic Szekeres-Szafron
metrics is mainly relevant from a conceptual point of view.
Nevertheless, in order to obtain physically realistic models there
are still many steps to take: (i)  obtain the complementary
constraints that the thermodynamic condition imposes on the metric
coordinate functions $\alpha$ and $\beta$; (ii) determine for these
thermodynamic models the expression $\chi(\rho, p)$ of the
indicatrix function; (iii) solve, for this indicatrix function, the
inverse problem of determining the thermodynamic scheme that defines
the thermodynamic properties of the fluid.

Starting from the conditions (\ref{lte-n}) Krasi\'nski {\em et al.}
\cite{KQS} proved that if a class I Szekeres-Szafron metric admits a
thermodynamic scheme then, necessarily, it admits symmetries.
Nevertheless, there are thermodynamic Szekeres-Szafron solutions of
class II without symmetries \cite{KQS}. The result of Krasi\'nski
{\em et al.} \cite{KQS} concerning class I can be easily found
following Eq. (\ref{lte-theta}). Indeed, if we impose Eq.
(\ref{lte-theta}) taking into account Eqs. ({\ref{density-II}) and
(\ref{expansion-II}), we obtain:
\begin{equation} \label{lte-Q}
 \dot{Q} F,_z h(d \nu)= 0 \, , \qquad F \equiv \frac{\ddot{\phi}}{\phi} \, ,
\end{equation}
where $Q$ is given in Eq. (\ref{Q}). The constraint $F,_z=0$ leads
to the FLRW limit. Then, from condition (\ref{lte-Q}), lemma
\ref{lemma-G3} and lemma \ref{lemma-Q} in Appendix \ref{app-B}, we
recover the Krasi\'nski {\em et al.} result \cite{KQS} (see also
Ref. \cite{Krasinski-Plebanski}):
\begin{proposition} \label{prop-termo-b}
A strict Szekeres-Szafron solution of type I represents a perfect
fluid in local thermal equilibrium if, and only if, the spacetime
admits a group of isometries $G_3$ on orbits $S_2$, that is, the
metric satisfies any of the equivalent conditions in proposition
{\em \ref{propo-G3}}.
\end{proposition}

All the above-quoted results contribute to the first step (i) in
looking for realistic models. Years ago we presented some
preliminary results on the inverse problem for type II sSS metrics
\cite{cfERE}. The exhaustive analyses of this subject, which require
the results of the recent paper \cite{fluperLTE}, is an ongoing work
that will be considered elsewhere.

\section{\label{sec-discussion}Discussion}
\label{sec-dis}

Starting from the invariant characterization by Barnes and
Rowlingson \cite{Barnes-Row}, in this paper we have presented two
IDEAL (intrinsic, deductive, explicit and algorithmic)
characterizations of the Szekeres-Szafron universes. The first one
is of the lowest order (first derivatives in the Riemann tensor) and
involves both the Weyl and the Ricci tensors (Sec. \ref{sec-ideal}).
The second one constitutes a Rainich-like approach and it requires
second-order conditions solely in terms of the Ricci tensor (Sec.
\ref{sec-Rainich}).

It is worth remarking that the conditions that we found in the
above-cited characterizations involve algebraic and differential
concomitants of invariant vectors, like the velocity $u$ and the
simple Weyl eigenvector $b$. These invariant vectors can be
explicitly obtained in terms of the associated projectors, which are
concomitants of the Riemann tensor. For example, $u =
-[-\Gamma(x,x)]^{-\frac12} \Gamma(x)$, where $\Gamma$ is given in
(\ref{fluper-definitions-A}) and $x$ is any time-like vector. Then,
when we impose any condition on $u$, this condition involves an
arbitrary vector $x$. In order to prevent this $x$ from appearing in
the characterization equations, we have opted to use concomitants of
the projector  $\Gamma$. Thus, for example, we worked with $\Sigma$
and $\Theta$ instead of with $\sigma$ and $\theta$. A similar
situation occurs with the simple Weyl eigenvector $b$ and its
associated projector $B$.

We have also explicitly labeled some significant subfamilies of the
Szekeres-Szafron solutions:    the sSS metrics of class I and class
II, which appear in a natural way when integrating the field
equations (Sec.  \ref{sec-beta}), and    the Szekeres-Szafron
metrics admitting a three-dimensional group of isometries $G_3$ on
space-like two-dimensional orbits $S_2$ (Sec.  \ref{sec-G3}). It is
worth remarking that the metrics of this last class are the geodesic
perfect fluid solutions with these symmetries (proposition
\ref{propo-G3-geodesic}). Moreover, there are several equivalent
invariant conditions labeling this class, and all of them impose
that the projection of an invariant vector on the space-like
principal plane (tangent to the orbits group) vanishes (propositions
\ref{propo-G3}, \ref{propo-h-theta}, \ref{propo-h-omega} and
\ref{propo-h-rho}).

The analysis of the subfamilies of the Szekeres-Szafron metrics
defined by first-order invariant conditions (linear in the first
derivatives) shows that a few significant classes can be considered
(Sec.  \ref{sec-class}). We have the type I and type II sSS metrics,
and those admitting a $G_3$ on $S_2$, which we have quoted in the
paragraph above as well as the solutions with constant pressure,
which generalize the dust solutions by Szekeres. The other defined
subfamilies necessarily admit symmetries: either a $G_3$ on $S_2$,
or a $G_1$ with the simple Weyl eigenvector $b$ tangent to the
orbits.

We have also given an IDEAL labeling of the Szekeres-Szafron
solutions that can be interpreted as a perfect fluid in local
thermal equilibrium (Sec.  \ref{sec-termo}).

It is worth noting the powerful   advantages of an IDEAL
characterization over the previous known invariant ones. Indeed, the
deductive, explicit and algorithmic qualities allow us to built a
flow chart with a finite number of steps and to implement it easily
by using the current tensor calculus packages.

The diagram below presents one of all possible flow charts that can
be built from our results. It allows us to distinguish the perfect
fluid solutions, the Szekeres-Szafron metrics (SS), the FLRW
universes and the strict Szekeres-Szafron metrics of classes I
(sSS-I) and II (sSS-II). We make use of the Ricci and Weyl
concomitants $\Gamma$, $s$, ${\cal A}$, $\Sigma$, $H$, $E$, $I$, $J$
and $B$ defined in Eqs. (\ref{fluper-definitions-A}),
(\ref{fluper-definitions-B}), (\ref{a-w}), (\ref{sigma}),
(\ref{H=0}), (\ref{E}) and (\ref{B}). These Riemann invariants can
be computed when they are involved in the equations of a specific
step (tags with outgoing arrows from the top and the left of the
diagram). The explicit conditions labelling the different families
of metrics in each step are presented inside diamonds. The labeled
families of metrics are reported inside rectangles.

\vspace{2mm}

%%%%%%%%%%%%%%%%%%%%%%

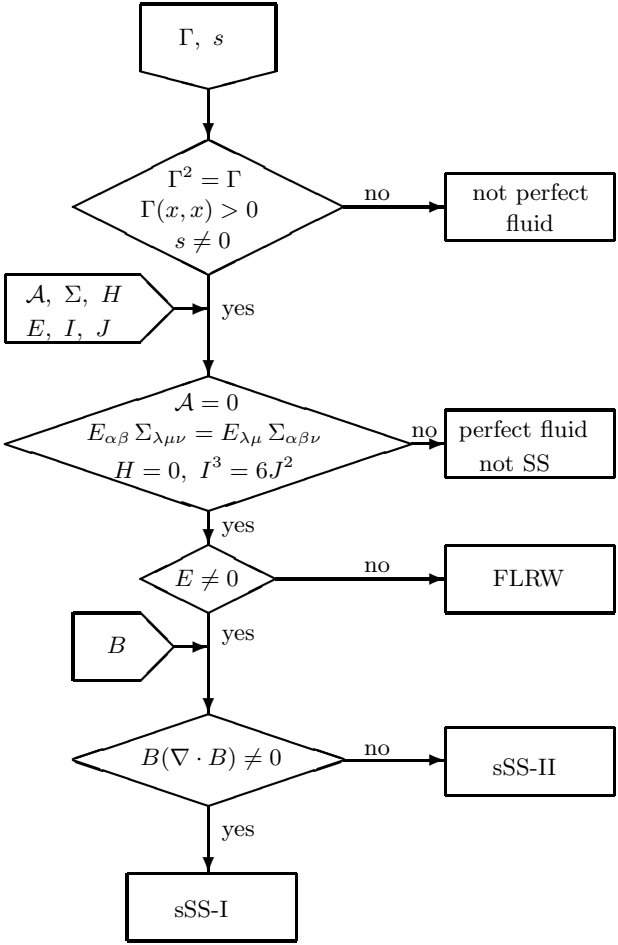
\setlength{\unitlength}{0.9cm} {\small \noindent
\begin{picture}(0,18)
\thicklines

\put(4,17){\line(-4,-1){1}}
 \put(2,17){\line(4,-1){1}}
\put(2,17){\line(0,1){1}} \put(4,18){\line(-1,0){2}}
\put(4,18){\line(0,-1){1}} \put(2.2,17.4){$ \ \ \ \Gamma , \ s   $}

%%%%%%%%%%%%%%%%
\put(3,16.75){\vector(0,-1){0.75}}

%%%%%%%%%%%%%%%
\put(1,15){\line(2,1){2}} \put(1,15){\line(2,-1){2}}
\put(5,15){\line(-2,1){2}}\put(5,15){\line(-2,-1){2}}

\put(2,14.85){$\Gamma (x,x) > 0 $}
 \put(2.4,15.3){$\Gamma^2 = \Gamma $}
\put(2.5,14.4){$s \neq 0 $}

%%%%%%%%%%%%%%%%%55
\put(5,15){\vector(1,0){1.5}} \put(5.3,15.1){no}
\put(3,14){\vector(0,-1){1.5}} \put(3.2,13.4){yes}

%%%%%%%%%%%%5

\put(6.5,14.5){\line(1,0){2.5}} \put(6.5,14.5){\line(0,1){1}}
\put(9,15.5){\line(-1,0){2.5}} \put(9 ,15.5){\line(0,-1){1}}

\put(6.9,15.1){{\rm not perfect}}
 \put(7.4,14.65){{\rm fluid}}
 %%%%%%%%%%%%%%%%

\put(0,13){\line(1,0){2}} \put(0,13){\line(0,1){1}}
\put(2,14){\line(-1,0){2}}

\put(2,14){\line(1,-1){0.5}}
 \put(2,13){\line(1,1){0.5}}

%%%%%%%%%%%%%%%%%%%%%%%

\put(2.5,13.5){\vector(1,0){0.5}}
%\put(2,14){\line(0,-1){1}}

\put(0.3,13.6){${\cal A}, \ \Sigma, \ H $}

\put(0.3,13.1){$E, \ I, \ J $}

%%%%%%%%%%%%%%%%%%%%%%%%5

\put(0,11.5){\line(3,1){3}}

\put(0,11.5){\line(3,-1){3}}

 \put(6,11.5){\line(-3,1){3}}
\put(6,11.5){\line(-3,-1){3}}
%\put(1,12){\line(2,1){2}} \put(1,12){\line(2,-1){2}}
%\put(5,12){\line(-2,1){2}} \put(5,12){\line(-2,-1){2}}

\put(2.5,12){${\cal A} = 0 $}

\put(1.6,11){$H=0, \ I^3 = 6 J^2 $}

\put(1.2,11.55){$E_{\alpha \beta}\,  \Sigma_{\lambda \mu \nu} =
E_{\lambda \mu}\, \Sigma_{\alpha \beta \nu}$}

%%%%%%%%%%%%%%%%%

\put(6.5,11){\line(1,0){2.5}} \put(6.5,11){\line(0,1){1}}
\put(9,12){\line(-1,0){2.5}} \put(9,12){\line(0,-1){1}}

\put(6.7,11.6){{\rm  perfect fluid}}
 \put(7,11.1){{\rm not SS}}

%%%%%%%%%%%%%%%%%

\put(6,11.5){\vector(1,0){0.5}} \put(6,11.6){no}

%%%%%%%%%%%%%%%%

\put(3,10){\line(-2,-1){1}} \put(3,10){\line(2,-1){1}}
\put(3,9){\line(2,1){1}} \put(3,9){\line(-2,1){1}}
\put(2.5,9.4){$E \neq 0$}
%\put(2.2,8.9){$B(\nabla B)\neq 0$}
 %%%%%%%%%%%%%%%

\put(3,10.5){\vector(0,-1){0.5}} \put(3.2,10.2){yes}

%%%%%%%%%%%%%%%%%
\put(4,9.5){\vector(1,0){2.5}}
 \put(5.3,9.6){no}
\put(3,9){\vector(0,-1){1.5}} \put(3.2,8.6){yes}

\put(6.5,9){\line(1,0){2.5}} \put(6.5,9){\line(0,1){1}}
\put(9,10){\line(-1,0){2.5}} \put(9,10){\line(0,-1){1}}

\put(7.2,9.4){{\rm  FLRW}}

%%%%%%%%%%%%%%

\put(1,8){\line(1,0){1}} \put(1,8){\line(0,1){1}}
\put(2,9){\line(-1,0){1}}

\put(2,9){\line(1,-1){0.5}}
 \put(2,8){\line(1,1){0.5}}

\put(2.5,8.5){\vector(1,0){0.5}}
 \put(1.5,8.4){$B$}

%%%%%%%%%%%

\put(3,7.5){\line(-3,-1){2}} \put(3,7.5){\line(3,-1){2}}
\put(3,6.15){\line(3,1){2}} \put(3,6.15){\line(-3,1){2}}
\put(2,6.75){$B(\nabla \cdot B)\neq 0$}

%%%%%%%%%%%%%5

\put(5,6.82){\vector(1,0){1.5}}
% \put(5.3,9.6){no}
%\put(3,9){\vector(0,-1){1.5}} \put(3.2,8.6){yes}

\put(6.5,6.3){\line(1,0){2.5}} \put(6.5,6.3){\line(0,1){1}}
\put(9,7.3){\line(-1,0){2.5}} \put(9,7.3){\line(0,-1){1}}

\put(7.2,6.6){{\rm  sSS-II}} \put(5.3,6.9){no}

\put(3,6.15){\vector(0,-1){1}} \put(3.2,5.7){yes}

%%%%%%%%%%%%%%%%%
\put(1.8,4.15){\line(1,0){2.5}} \put(1.8,4.15){\line(0,1){1}}
\put(4.3,5.15){\line(-1,0){2.5}} \put(4.3,5.15){\line(0,-1){1}}
\put(2.5,4.5){{\rm  sSS-I}}

\end{picture}
}

%%%%%%%%%%%%%%%%%%%%%%%
\vspace{-3.5cm}

As commented in the Introduction, our IDEAL approach is an
alternative to the Cartan-Brans-Karlhede method to analyze the
equivalence of two metric tensors. This method is based on working
in an orthonormal (or a null) frame fixed by the underlying geometry
of the Riemann tensor. Nevertheless, the historic theorems that
characterize locally flat Riemann spaces, Riemann spaces with a
maximal group of isometries, and locally conformally flat Riemann
spaces show that the determination of a Riemannian canonical frame
is not necessary in labeling specific families of spacetimes. The
conditions applied in these theorems involve explicit concomitants
of the curvature tensor (Riemann, Weyl and Cotton tensors) and,
consequently, they are IDEAL characterizations. We find a similar
situation in characterizing other physically relevant families of
spacetimes, such as the Stephani and the FLRW universes.

A suitable procedure is to analyze every particular case in order to
understand the minimal set of elements of the curvature tensor that
are necessary to label these geometries, an approach adapted to each
particular geometry we want to characterize. This is the method we
have achieved here in labeling the Szekeres-Szafron metrics, and it
is also the one used in previous articles when characterizing
different families of solutions \cite{fs-SSST, fs-ssst-Ricci, fsI,
fsS, fms, fsWEM, fsKY, fsIa, fsD, fsWEM2, cfs, fsIb, fs-RainichEM-D,
fsEM-sym, fsKerr, fswarped, fs-typeD}.

\begin{acknowledgments}
This work has been supported by the Spanish ``Ministerio de
Econom\'{\i}a y Competitividad", MINECO-FEDER project
FIS2015-64552-P.
\end{acknowledgments}

\appendix

\section{\label{app-A}Szekeres-Szafron metrics: coordinate functions and invariant scalars.}

The specific form of the coordinate functions $\alpha(t,z,x,y)$ and
$\beta(t,z,x,y)$, and the expression of the pressure and energy
density for Szekeres-Szafron metrics of classes I and II can be
found in several papers \cite{Krasinski} \cite{Krasinski-Plebanski}
\cite{Szafron}. Now we give these expressions by using the notation
in Ref. \cite{Krasinski-Plebanski} with a few changes. We also offer
expressions for the simple Weyl eigenvalue and the expansion of the
fluid.
\\[3mm]
{\bf Szekeres-Szafron metrics of class I ($\beta,_z \not=0$)}\\[2mm]
{\em Coordinate functions}:
\begin{equation}\label{I-alpha-beta}
e^{\alpha} = \phi,_z +  \phi  \nu,_z  , \quad  e^{\beta} =\phi S^{-1}
 , \quad   \phi = \phi(t,z),
\end{equation}
{\small
\begin{equation}
{\rm S}(z,x,y) \equiv  \frac12 U(z) (x^2 + y^2) + V_1 (z) x +  V_2
(z) y + 2\,W(z).  \,   \label{I-S}
\end{equation}
}
{\em Pressure and energy density}:
{\small
\begin{equation}
p =  - \! \left[\frac{2 \ddot{\phi}}{\phi} + \frac{\dot{\phi}^2}{\phi^2}
+ \frac{k(z)}{\phi^2}\right] \! , \,  \ k(z) \equiv 4 UW\! - \!V_1^2 \! - \!V_2^2
\! - \! 1, \label{pressure-I}
\end{equation}}
\begin{equation}
  \rho = \frac{k'  + (\dot{\phi}^2+k)[(\ln \phi),_z + 3 \nu,_z]
+ 2 \dot{\phi} \, \dot{\phi},_{z}}{ \phi (\phi,_z + \, \phi
\, \nu,_z)} \,  . \label{density-I}
\end{equation}
%
%\ \\
%
{\em Simple Weyl eigenvalue}:
\begin{eqnarray} \label{eigenvalue-I}
\omega = \frac{\phi \, \ddot{\phi},_z -\, \ddot{\phi}\, \phi,_z}{3
\phi (\phi,_z + \, \phi \, \nu,_z)}   \,  . \label{density-I}
\end{eqnarray}
{\em Expansion}:
\begin{eqnarray} \label{expansion-I}
\theta = \frac{\dot{\phi},_z + \, \dot{\phi} \, \nu,_z}{\phi,_z + \,
\phi \, \nu,_z} + \frac{2 \dot{\phi}}{ \phi}  \,  .
\label{density-I}
\end{eqnarray}
\ \\[0mm]
{\bf Szekeres-Szafron metrics of class II ($\beta,_z =0$)}\\[2mm]
{\em Coordinate functions}:
\begin{eqnarray} \label{II-alpha-beta}
e^{\alpha} = \lambda + \phi P  , \quad  e^{\beta} = \phi \, C  ,\quad   P={\rm S} \,C , \quad \\[2mm]
\phi = \phi(t), \quad \lambda=\lambda(t,z), \quad {\rm S}= {\rm
S}(z,x,y), \quad  \\[1mm]
 C(x,y) \equiv [1 + \frac{k}{4} (x^2 + y^2)]^{-1}, \quad k \equiv 0, 1, -1  , \quad \label{II-C-S}
\end{eqnarray}
\begin{equation}
 \phi \ddot{\lambda} + \dot{\lambda} \dot{\phi} -
\lambda  \frac{\phi \ddot{\phi} + \dot{\phi}^2 + k}{\phi} = U+kW \, .
\label{II-phi-S}
\end{equation}
{\em Pressure and energy density}:
\begin{eqnarray} \label{pressure-II}
p = -\left[\frac{2 \ddot{\phi}}{\phi} + \frac{\dot{\phi}^2}{\phi^2} + \frac{k}{\phi^2}\right]   \,  ,  \\[1mm]
\rho = \frac{2 ( \ddot{\phi} \, \lambda - \phi \,
\ddot{\lambda})}{\phi (\lambda + \, \phi \, P)} + \frac{3
\dot{\phi^2}}{ \phi^2} + \frac{3k}{\phi^2} \,  . \label{density-II}
\end{eqnarray}
{\em Simple Weyl eigenvalue}:
\begin{equation} \label{eigenvalue-II}
\omega = \frac{\phi \, \ddot{\lambda} -\, \ddot{\phi} \, \lambda}{3
\phi (\lambda + \, \phi \, P)}   \,  .
\end{equation}
{\em Expansion}:
\begin{equation} \label{expansion-II}
\theta = \frac{\dot{\lambda} + \dot{\phi} \,P }{\lambda + \, \phi \,
P} + \frac{2 \dot{\phi}}{ \phi}  \,  .
\end{equation}
%

%%%%%%%%%%%%%%%%%%%%%%%%

\section{\label{app-B}Some lemmas}

\begin{lemma} \label{lemma-omega-r-I}
For a strict Szekeres-Szafron metric of class I, the condition
$\omega,_z =0$ implies $\nu,_{zx} = \nu,_{zy} =0$.
\end{lemma}
{\em Proof:} Let us  suppose that $\nu,_{zx} \neq 0$. From the
expression (\ref{eigenvalue-I}) for $\omega$ we have that
{\small
 \be \label{B1}
\hspace{-1.5mm} \omega,_{x} =  \frac{- \omega \, \phi
\nu,_{zx}}{\phi,_{z} + \phi \nu,_{z}}  , \ \  \omega,_{zx} = \frac{-
\omega,_{z} \, \phi \nu,_{zx}}{\phi,_{z} + \phi \nu,_{z}} - \omega
\left[ \frac{\phi \nu,_{zx}}{\phi,_{z} + \phi \nu,_{z}} \right]_{,z} \! \!
. \ee }
Furthermore, a straightforward calculation gives
\be \label{B2} \left[ \frac{\phi \nu,_{zx}}{\phi,_{z} + \phi
\nu,_{z}} \right]_{,x} =  \frac{\phi \nu,_{zx}}{\phi,_{z} + \phi
\nu,_{z}} \left[ \frac{\nu,_{rxx}}{\nu,_{rx}} -   \frac{\phi
\nu,_{zx}}{\phi,_{z} + \phi \nu,_{z}} \right] . \ee
If $\omega_{,z }=0$, the expression (\ref{B1}) implies that $\left[
\frac{\phi \nu_{,zx}}{\phi_{,z} + \phi \nu,_{z}} \right]_{,z} =0$.
Then, by taking the derivative with respect to $z$ in Eq. (\ref{B2})
we get that $[\ln \nu,_{zx}],_{xz} = 0$. If we compute this equation
by taking into account the expression of $\nu$ in Eq. (\ref{I-S}),
and equate the coefficients in the powers of $x$ and $y$, we get
that $\nu_{,zx}=0$. Similarly we have that $\nu_{,zy}=0$.

%%%%%

\begin{lemma} \label{lemma-omega-r-II}
For a strict Szekeres-Szafron metric of class II, the condition
$\omega,_z =0$ implies,
\begin{eqnarray}
e^{\alpha}  = \lambda (t) + \phi(t) {\rm S}(x, y) \, , \\[1mm]   {\rm
S}(x, y) \equiv \frac12 (x^2 + y^2) + c_1 x + c_2 y + 2c \, ,
\end{eqnarray}
where $c_i$ and $c$ are constants. Consequently $\xi = \partial_z$
is a Killing vector.
\end{lemma}
{\em Proof:} Without loss of generality we can take $W(z)=0$ in the
expression (\ref{I-S}) of the function $S$ of a metric of class II
\cite{Krasinski}. From the expression (\ref{eigenvalue-II}) of
$\omega$, and taking into account that $\phi\equiv \phi(t)$, it
holds that $\omega,_{z}=0$ is equivalent to $(\phi \omega),_{z}=0$.
A straightforward calculation shows that this condition can be
stated as
%
%{\small
$$ G,_{z}    [ \lambda C^{-1}  + \phi S ] = G \ [
\lambda,_{z} C^{-1} + \phi S,_{z} ], \ \ G \equiv  \phi
\ddot{\lambda} - \ddot{\phi} \lambda.  $$
%}
%
If we expand the expression above and equate the coefficients  in
the different powers of $x$ and $y$, we have that
$$\frac{G_{,z}}{G} = \frac{\lambda_{,z}}{\lambda} =
\frac{U^{\prime}}{U} = \frac{V_1^{\prime}}{V_{1}} =
\frac{V_2^{\prime}}{V_2} \, .$$
Integrating these equations we get $(\lambda/U),_z = (P/U),_z = 0$.
Thus, a redefinition of the coordinate $z$ leads to the lemma.

%%%%%%%

\begin{lemma} \label{lemma-I-rhoz}
For a Szekeres-Szafron metric of class I, the condition $\rho,_z =0$
leads necessarily to a FLRW universe.
\end{lemma}
{\em Proof:} If $\rho,_z=0$, then we have $\dot{\rho},_{z} =0$ and,
from the conservation condition, we get that $\theta_{,z}=0$. On the
other hand, from  Eq. (\ref{rho-omega}), we get that $\rho_{,z}=0$
is equivalent to $\omega_{,z} + [\ddot{\phi}/\phi],_{z} =0 $.  If we
expand this last condition, and use $\dot{\theta},_{z}=0$ to replace
$\ddot{\phi},_{zz}$, we obtain that $\rho_{,z}=0$ implies $\phi
\dot{\phi}_{,z} - \dot{\phi} \phi_{,z} $. Then, from Eq.
(\ref{eigenvalue-I}), $\omega=0$ and the solution is a FLRW
universe.

%%%%%%

\begin{lemma} \label{lemma-Q}
For a strict Szekeres-Szafron metric of type I, condition $\dot{Q}
=0$ implies $\nu,_{zx} = \nu,_{zy} =0$, where

\begin{eqnarray}
\hspace{-22mm} Q=   - \nu,_{z} \left[ \frac{F,_{zz}}{F,_{z}} + 6 \,
\frac{\phi,_{z}}{\phi} \right] - 3 \nu,_{z} \nu,_{z} + \qquad \qquad  \nonumber \\
  + \left[ \frac{\phi,_{zz}}{\phi} - \frac{\phi,_{z}}{\phi} \
\frac{F,_{zz}}{F,_{z}} - 4 \frac{\phi,_{z} \phi,_{z}}{\phi^2}
\right]   , \quad F \equiv \frac{\ddot{\phi}}{\phi} \, .\label{Q}
\end{eqnarray}
\end{lemma}
{\em Proof:}  The condition $\dot{Q} =0$, implies $h
(\dif{\dot{Q}})$. If we expand this last expression we get:
$$\left[ \frac{F_{,zz}}{F_{,z}} + 6 \, \frac{\phi_{,z}}{\phi}
\right]_{,t} \ h(\dif{\nu_{,z}}) \!=0 \, .$$
Thus, either $h(\dif{\nu_{,z}}) =0$ or $\left[
\frac{F_{,zz}}{F_{,z}} + 6 \, \frac{\phi_{,z}}{\phi} \right]_{,t}
=0$.
 Then, if $h (\dif \nu_{,z})\neq 0$, the condition $\dot{Q}=0$ is equivalent
to the two equations
$$ \left[ \frac{F_{,zz}}{F_{,z}} + 6 \, \frac{\phi_{,z}}{\phi}
\right]_{,t} \! \!=0, \ \  \left[ \frac{\phi_{,zz}}{\phi} -
\frac{\phi_{,z}}{\phi} \ \frac{F_{,zz}}{F_{,z}} - 4 \frac{\phi_{,z}
\phi_{,z}}{\phi^2} \right]_{,t} \!\!=0 . $$
The first equation can be partially integrated and we have that two
functions $a(z)$ and $b(t)$  exist such that
$$F_{,z} \phi^6 = e^{a(z)} e^{b(t)} .$$
Putting this into the second equation, we have that the function
$\phi$ necessarily factorizes, and we obtain a FLRW universe.

\nocite{*}
\bibliography{Szafron-ideal-preprint}% Produces the bibliography via BibTeX.

\end{document}